\documentclass[11pt]{article}

\usepackage{amssymb,amsmath}
\usepackage{amsthm}
\usepackage{amsfonts}
\usepackage{latexsym}
\usepackage{geometry}
\usepackage{graphicx}
\usepackage{enumerate}
\usepackage{color}
\usepackage{cite}
\usepackage{tikz}

\usepackage[latin1]{inputenc}

\newtheorem{thm}{Theorem}
\newtheorem{cor}[thm]{Corollary}

\newtheorem{lem}[thm]{Lemma}

\theoremstyle{definition}
\newtheorem{defn}{Definition}
\newtheorem{exa}{Example}

\newcommand{\gp}[2]{\ensuremath{#1^{#2}}}
\newcommand{\gpp}[2]{\ensuremath{#1^{\mathcal{N}(#2)}}}
\newcommand{\gme}[2]{\ensuremath{#1^{#2}}}
\newcommand{\Path}[1]{\ensuremath{{\sf P}_{#1}}}
\newcommand{\Star}[1]{\ensuremath{{\sf S}_{#1}}}
\newcommand{\C}[1]{\ensuremath{{\sf C}_{#1}}}
\newcommand{\K}[1]{\ensuremath{{\sf K}_{#1}}}
\newcommand{\CS}[2]{\ensuremath{{\sf CS}_{#1,#2}}}
\newcommand{\tur}[2]{\ensuremath{{\sf T}_{#1,#2}}}
\newcommand{\turc}[2]{\ensuremath{{\sf TC}_{#1,#2}}}
\newcommand{\Dec}[4]{\ensuremath{{\mathcal D}(#1,#2,#3,#4)}}
\newcommand{\F}{F}
\newcommand{\al}{\alpha}
\newcommand{\fT}{f_{\sf T}}
\newcommand{\fTC}{f_{\sf TC}}

\makeatletter  

\title{Tur\'an Graphs, Stability Number, and Fibonacci Index}
\author{V\'eronique Bruy\`ere\footnote{Department of Theoretical Computer Science, Universit\'e de  Mons-Hainaut, Avenue du Champ de Mars 6, B-7000 Mons, Belgium.} \and 
Hadrien M\'elot\footnotemark[1] $^{,}$\footnote{Charg\'e de Recherches F.R.S.-FNRS. Corresponding author. E-mail: {\tt hadrien.melot@umh.ac.be}.}
}
\date{22 February, 2008}

\begin{document}

\maketitle
\vspace*{0.2cm}

\hrule
\vspace*{0.2cm}
\small
\noindent
\textbf{Abstract.} The Fibonacci index of a graph is the number of its stable sets. This parameter is widely studied and has applications in chemical graph theory. In this paper, we establish tight upper bounds for the Fibonacci index in terms of the stability number and the order of general graphs and connected graphs. Tur\'an graphs frequently appear in extremal graph theory. We show that Tur\'an graphs and a connected variant of them are also extremal for these particular problems. 

\vspace*{0.2cm}
\noindent
\emph{Keywords:} Stable sets; Fibonacci index; Merrifield-Simmons index; Tur\'an graph; $\al$-critical graph.

\vspace*{0.2cm}
\hrule

\normalsize

\section{Introduction}

The Fibonacci index $\F(G)$ of a graph $G$ was introduced in 1982 by Prodinger and Tichy~\cite{PT82} as the number of stable sets in $G$. In 1989, Merrifield and Simmons~\cite{MS89} introduced independently this parameter in the chemistry literature\footnote{The Fibonacci index is called the Fibonacci number by Prodinger and Tichy~\cite{PT82}.  Merrifield and Simmons introduced it as the $\sigma$-index~\cite{MS89}, also known as the Merrifield-Simmons index.}. They showed that there exist correlations between the boiling point and the Fibonacci index of a molecular graph. Since, the Fibonacci index has been widely studied, especially during the last few years. The majority of these recent results appeared in chemical graph theory~\cite{LLW03, LZG05, TW05, Wag07, WH06, WHW07} and in extremal graph theory~\cite{HW07, KTWZ07, PV, PV05, PV06}. 

In this literature, several results are bounds for $\F(G)$ among graphs in particular classes. Lower and upper bounds inside the classes of general graphs, connected graphs, and trees are well known (see Section~\ref{sec_prelim}).  Several authors give a characterization of trees with maximum Fibonacci index inside the class ${\cal T}(n,k)$ of trees with order $n$ and a fixed parameter $k$. For example, Li et al.~\cite{LZG05} determine such trees when $k$ is the diameter; Heuberger and Wagner~\cite{HW07} when $k$ is the maximum degree; and Wang et al.~\cite{WHW07} when $k$ is the number of pending vertices. Unicyclic graphs are also investigated in similar ways~\cite{PV05,PV06,WH06}. 

The Fibonacci index and the stability number of a graph are both related to stable sets. Hence, it is natural to use the stability number as a parameter to determine bounds for $\F(G)$. Let ${\cal G}(n,\al)$ and ${\cal C}(n,\al)$ be the classes of -- respectively general and connected -- graphs with order $n$ and stability number $\al$. The lower bound for the Fibonacci index is known for graphs in these classes. Indeed, Pedersen and Vestergaard~\cite{PV06} give a simple proof to show that if $G \in {\cal G}(n,\al)$ or $G \in {\cal C}(n,\al)$, then $\F(G) \ge 2^\al + n - \al$. Equality occurs if and only if $G$ is a complete split graph (see Section~\ref{sec_prelim}). In this article, we determine upper bounds for $\F(G)$ in the classes ${\cal G}(n,\al)$ and ${\cal C}(n,\al)$. In both cases, the bound is tight for every possible value of $\al$ and $n$ and the extremal graphs are characterized.

A Tur\'an graph is the union of disjoint balanced cliques. Tur\'an graphs frequently appear in extremal graph theory. For example, the well-known Theorem of Tur\'an~\cite{Turan} states that these graphs have minimum size inside ${\cal G}(n,\al)$. We show in Section~\ref{sec_graph} that Tur\'an graphs have also  maximum Fibonacci index inside ${\cal G}(n,\al)$. Observe that removing an edge in a graph strictly increases its Fibonacci index. Indeed, all existing stable sets remain and there is at least one more new stable set: the two vertices incident to the deleted edge. Therefore, we might have the intuition that the upper bound for $\F(G)$ is a simple consequence of the Theorem of Tur\'an. However, we show that it is not true (see Sections~\ref{sec_prelim} and \ref{sec_conc}). The proof uses structural properties of $\al$-critical graphs.

Graphs in ${\cal C}(n,\al)$ which maximize $\F(G)$ are characterized in Section~\ref{sec_conn}. We call them Tur\'an-connected graphs since they are a connected variant of Tur\'an graphs. It is interesting to note that these graphs again minimize the size inside ${\cal C}(n,\al)$. Hence, our results lead to questions about the relations between the Fibonacci index, the stability number, the size and the order of graphs. These questions are summarized in Section~\ref{sec_conc}.

\section{Basic properties} \label{sec_prelim}

In this section, we suppose that the reader is familiar with usual notions of graph theory (we refer to Berge~\cite{Berge01} for more details). First, we fix our terminology and notation. We then recall the notion of $\al$-critical graphs and give properties of such graphs, used in the next sections. We end with some basic properties of the Fibonacci index of a graph.

\subsection{Notations}

Let $G=(V,E)$ be a simple and undirected graph order $n(G) = | V |$ and size $m(G) = |E|$.  For a vertex $v \in V(G)$, we denote by $N(v)$ the neighborhood of $v$; its closed neighborhood is defined as $\mathcal{N}(v) = N(v) \cup \{v\}$. The degree of a vertex $v$ is denoted by $d(v)$ and the maximum degree of~$G$ by $\Delta(G)$. We use notation $G \simeq H$ when $G$ and $H$ are isomorphic graphs. The complement of $G$ is denoted by $\overline{G}$.

The \emph{stability number} $\al(G)$ of a graph $G$ is the number of vertices of a maximum stable set of $G$. Clearly, $1 \leq \al(G) \leq n(G)$, and $1 \leq \al(G) \leq n(G)-1$ when $G$ is connected.

\begin{defn}
We denote by $\gp{G}{v}$ the induced subgraph obtained by removing a vertex $v$ from a graph $G$. Similarly, the graph $\gpp{G}{v}$ is the induced subgraph obtained by removing the closed neighborhood of $v$. Finally, the graph obtained by removing an edge $e$ from $G$ is denoted by $\gme{G}{e}$.
\end{defn}

\begin{figure}[!ht]
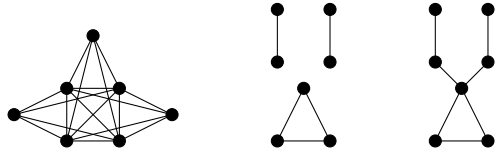

\begin{center}
\tikzstyle{every node}=[circle, draw, fill=black,
                        inner sep=0mm, minimum width=1.6mm]
\tikz[scale=0.7] {
\draw (1,0) node {} -- (2,0) node {}; 
\draw (1,1) node {} -- (2,1) node {};
\draw (1,0) -- (1,1); 
\draw (2,0) -- (2,1); 
\draw (1,0) -- (2,1);
\draw (2,0) -- (1,1);
\draw (0,0.5) node {} -- (1,0);
\draw (0,0.5) -- (2,0);
\draw (0,0.5) -- (1,1);
\draw (0,0.5) -- (2,1);
\draw (3,0.5) node {} -- (1,0);
\draw (3,0.5) -- (2,0);
\draw (3,0.5) -- (1,1);
\draw (3,0.5) -- (2,1);
\draw (1.5,2) node {} -- (1,0);
\draw (1.5,2) -- (2,0);
\draw (1.5,2) -- (1,1);
\draw (1.5,2) -- (2,1);
\draw (5,0) node {} -- (6,0) node {}; 
\draw  (5.5,1) node {}  -- (5,0);
\draw  (5.5,1) -- (6,0);
\draw (5,1.5) node {} -- (5,2.5) node {};
\draw (6,1.5) node {} -- (6,2.5) node {}; 
\draw (8,0) node {} -- (9,0) node {}; 
\draw  (8.5,1) node {}  -- (8,0);
\draw  (8.5,1) -- (9,0);
\draw (8,1.5) node {} -- (8,2.5) node {};
\draw (9,1.5) node {} -- (9,2.5) node {}; 
\draw  (8.5,1) -- (9,1.5);
\draw  (8.5,1) -- (8,1.5);
}                        
\caption{The graphs $\CS{7}{3}$, $\tur{7}{3}$ and $\turc{7}{3}$} \label{fig_exagr}
\end{center}
\end{figure}

Classical graphs of order $n$ are used in this article: the complete graph $\K{n}$, the path $\Path{n}$, the cycle $\C{n}$, the star $\Star{n}$ (composed by one vertex adjacent to $n-1$ vertices of degree 1) and the complete split graph $\CS{n}{\al}$ (composed of a stable set of $\al$ vertices, a clique of $n-\al$ vertices and each vertex of the stable set is adjacent to each vertex of the clique). The complete split graph $\CS{7}{3}$ is depicted in Figure~\ref{fig_exagr}.

We also deeply study the two classes of Tur\'an graphs and Tur\'an-connected graphs. A \emph{Tur\'an graph} $\tur{n}{\al}$ is a graph of order $n$ and a stability number $\al$ such that $1 \leq \al \leq n$, that is defined as follows. It is the union of $\al$ disjoint balanced cliques (that is, such that their orders differ from at most one)~\cite{Turan}. These cliques have thus $\lceil \frac{n}{\al} \rceil$ or $\lfloor \frac{n}{\al} \rfloor$ vertices. We now define a \emph{Tur\'an-connected graph} $\turc{n}{\al}$ with $n$ vertices and a stability number $\al$ where $1 \leq \al \leq n-1$. It is constructed from the Tur\'an graph $\tur{n}{\al}$ with $\al-1$ additional edges. Let $v$ be a vertex of one clique of size $\lceil \frac{n}{\al} \rceil$, the additional edges link $v$ and one vertex of each remaining cliques. Note that, for each of the two classes of graphs defined above, there is only one graph with given values of $n$ and $\al$, up to isomorphism. 

\begin{exa} \label{exa_Turan}
Figure \ref{fig_exagr} shows the Tur\'an graph $\tur{7}{3}$ and the Tur\'an-connected graph $\turc{7}{3}$. When $\al = 1$, we observe that $\tur{n}{1} \simeq \turc{n}{1} \simeq  \CS{n}{1}  \simeq \K{n}$. When $\al = n$, we have $\tur{n}{n} \simeq \CS{n}{n} \simeq  \overline{\K{n}}$, and when $\al = n-1$, we have $\turc{n}{n-1} \simeq \CS{n}{n-1}  \simeq \Star{n}$.
\end{exa}

\subsection{$\al$-critical graphs}

We recall the notion of $\al$-critical graphs~\cite{ErdG61, Joret07, Lovasz86}. An edge $e$ of a graph $G$ is $\al$\emph{-critical} if $\al(\gme{G}{e}) > \al(G)$, otherwise it is called $\al$\emph{-safe}. A graph is said to be $\al$\emph{-critical} if all its edges are $\al$-critical. By convention, a graph with no edge is also $\al$-critical. These graphs play an important role in extremal graph theory~\cite{Joret07}, and also in our proofs.

\begin{exa} \label{exa_critical}
Simple examples of $\al$-critical graphs are complete graphs and odd cycles. Tur\'an graphs are also $\al$-critical. On the contrary, Tur\'an-connected graph are not $\al$-critical, except when $\al =  1$.
\end{exa}

We state some interesting properties of $\al$-critical graphs.

\begin{lem} \label{lem_crit1}
Let $G$ be an $\al$-critical graph. If $G$ is connected, then the graph $\gp{G}{v}$ is connected for all vertices $v$ of $G$.
\end{lem}

\begin{proof}
We use two known results on $\al$-critical graphs (see, e.g., \cite[Chapter 12]{Lovasz86}). If a vertex $v$ of an $\al$-critical graph has degree 1, then $v$ and its neighbor $w$ form a connected component of the graph. Every vertex of degree at least 2 in an $\al$-critical graph is contained in a cycle. 

Hence, by the first result, the minimum degree of $G$ equals 2, except if $G \simeq \K{2}$. Clearly $\gp{G}{v}$ is connected by the second result or when $G \simeq \K{2}$.
\end{proof}

\begin{lem} \label{lem_crit}
Let $G$ be an $\al$-critical graph. Let $v$ be any vertex of $G$ which is not isolated. Then,
\[
 \al(G) = \al(\gp{G}{v}) = \al(\gpp{G}{v}) + 1.
\]
\end{lem}

\begin{proof}
Let $e =vw$ be an edge of $G$ containing $v$. Then, there exist in $G$ two maximum stable sets $S$ and $S'$, such that $S$ contains $v$, but not $w$, and $S'$ contains $w$, but not $v$ (see, e.g., \cite[Chapter 12]{Lovasz86}). Thus, $\al(G) = \al(\gp{G}{v})$ due to the existence of $S'$. The set $S$ avoids each vertex of $N(v)$. Hence, $S \setminus \{v\}$ is a stable set of the graph $\gpp{G}{v}$ of size $\al(G) - 1$. It is easy to check that this stable set is maximum.
\end{proof}

\subsection{Fibonacci index}

Let us now recall the Fibonacci index of a graph~\cite{PT82, MS89}. The \emph{Fibonacci index} $\F(G)$ of  a graph~$G$ is the number of all the stable sets in $G$, including the empty set. The following lemma about $\F(G)$ is well-known (see \cite{GP86, LZG05, PT82}). It is used intensively through the article.

\begin{lem} \label{lem_main1}
Let $G$ be a graph. 
\begin{itemize}
\item Let $e$ be an edge of $G$, then $\F(G) < \F(\gme{G}{e})$.
\item Let $v$ be a vertex of $G$, then $\F(G) = \F(\gp{G}{v}) + \F(\gpp{G}{v})$.
\item
If $G$ is the union of $k$ disjoint graphs $G_i$, $1\leq i\leq k$, then $\F(G) = \prod_{i=1}^k \F(G_i)$.
\end{itemize}
\end{lem}

\begin{exa} \label{exa_F}
We have $\F(\K{n}) = n+1$, $\F(\overline{\K{n}}) = 2^n$, $\F(\Star{n}) = 2^{n-1} + 1$ and $\F(\Path{n}) = f_{n+2}$ (recall that the sequence of Fibonacci numbers $f_n$ is $f_0 = 0, f_1= 1$ and $f_n = f_{n-1} + f_{n-2}$  for $n > 1$).
\end{exa}

Prodinger and Tichy~\cite{PT82} give simple lower and upper bounds for the Fibonacci index. We recall these bounds in the next lemma. 
\begin{lem} \label{lem_main2}
Let $G$ be a graph of order $n$.
\begin{itemize}
\item Then $n+1 \le \F(G) \le 2^n$ with equality if and only if $G \simeq \K{n}$ (lower bound) and $G \simeq \overline{\K{n}}$ (upper bound).
\item If $G$ is connected, then $n+1 \le \F(G) \le 2^{n-1} + 1$ with equality if and only if $G \simeq \K{n}$ (lower bound) and $G \simeq \Star{n}$ (upper bound).
\item If $G$ is a tree, then $f_{n+2} \le \F(G) \le 2^{n-1} + 1$ with equality if and only if $G \simeq \Path{n}$ (lower bound) and $G \simeq \Star{n}$ (upper bound).
\end{itemize}
\end{lem}

We denote by ${\cal G}(n,\al)$ the class of general graphs with order $n$ and stability number $\al$; and by ${\cal C}(n,\al)$ the class of connected graphs with order $n$ and stability number $\al$. Pedersen and Vestergaard~\cite{PV06} characterize graphs with minimum Fibonacci index as indicated in the following theorem.
\begin{thm} \label{thm_lb}
Let $G$ be a graph inside ${\cal G}(n,\al)$ or ${\cal C}(n,\al)$, then 
\[
\F(G) \ge 2^\al + n - \al,
\]
with equality if and only if $G \simeq \CS{n}{\al}$.
\end{thm}
The aim of this article is the study of graphs with maximum Fibonacci index inside the two classes ${\cal G}(n,\al)$ and ${\cal C}(n,\al)$. The system GraPHedron~\cite{GphDesc} allows a formal framework to conjecture optimal relations among a set of graph invariants. Thanks to this system, graphs with maximum Fibonacci index inside each of the two previous classes have been computed for small values of $n$~\cite{GPHRepFiboAlpha}. We observe that these graphs are isomorphic to Tur\'an graphs for the class ${\cal G}(n,\al)$, and to Tur\'an-connected graphs for the class ${\cal C}(n,\al)$. For the class ${\cal C}(n,\al)$, there is one exception when $n = 5$ and $\al = 2$: both the cycle $\C{5}$ and the graph $\turc{5}{2}$ have maximum Fibonacci index. 

Recall that the classical Theorem of Tur\'an~\cite{Turan} states that Tur\'an graphs $\tur{n}{\al}$ have minimum size inside ${\cal G}(n,\al)$. We might think that Tur\'an graphs have maximum Fibonacci index inside ${\cal G}(n,\al)$ as a direct corollary of the Theorem of Tur\'an and Lemma~\ref{lem_main1}. This argument is not correct since removing an $\al$-critical edge increases the stability number. Therefore,  Lemma~\ref{lem_main1} only implies that graphs with maximum Fibonacci index inside ${\cal G}(n,\al)$ are $\al$-critical graphs. In Section~\ref{sec_conc}, we make further observations on the relations between the size and the Fibonacci index inside the classes ${\cal G}(n,\al)$ and ${\cal C}(n,\al)$.

There is another interesting property of Tur\'an graphs related to stable sets. Byskov~\cite{Byskov04} establish that Tur\'an graphs have maximum number of maximal stable sets inside ${\cal G}(n,\al)$. The Fibonacci index counts not only the maximal stable sets but all the stable sets. Hence, the fact that Tur\'an graphs maximize $F(G)$ cannot be simply derived from the result of Byskov.

\section{General graphs}  \label{sec_graph}

In this section, we study graphs with maximum Fibonacci index inside the class ${\cal G}(n,\al)$. These graphs are said to be \emph{extremal}. For fixed values of $n$ and $\al$, we show that there is one extremal graph up to isomorphism, the Tur\'an graph $\tur{n}{\al}$ (see Theorem~\ref{thm_genUB}). 

Before establishing this result, we need some auxiliary results. We denote by $\fT(n,\al)$ the Fibonacci index of the Tur\'an graph $\tur{n}{\al}$. By Lemma~\ref{lem_main1}, its value is equal to
\[
\fT(n,\al) = \left(\left\lceil\frac{n}{\al}\right\rceil + 1\right)^p \  \left(\left\lfloor\frac{n}{\al} \right\rfloor + 1 \right)^{\al-p},
\]
where $p = (n \mod \al)$. We have also the following inductive formula.

\begin{lem} \label{lem_ind}
Let $n$ and $\al$ be integers such that $1 \le \al \le n$. Then
\[
\fT(n,\al) = \left\{ 
\begin{array}{l l l}
n+1 & \textrm{if} & \al = 1,\\
2^n & \textrm{if} & \al = n,\\
\fT(n-1,\al) +  \fT(n-\left\lceil \frac{n}{\al}\right\rceil,\al -1) & \textrm{if} & 2 \le \al \le n-1.\\
\end{array}
\right.
\]
\end{lem}

\begin{proof}
The cases $\al = 1$ and $\al = n$ are trivial (see Example~\ref{exa_F}). Suppose $2 \le \al \le n-1$. Let $v$ be a vertex of $\tur{n}{\al}$ with maximum degree. Thus $v$ is in a $\left\lceil\frac{n}{\al}\right\rceil$-clique. As $\al < n$, the vertex $v$ is not isolated. Therefore $\gp{\tur{n}{\al}}{v} \simeq \tur{n-1}{\al}$. As $\al \geq 2$, the graph $\gpp{\tur{n}{\al}}{v}$ has at least one vertex, and $\gpp{\tur{n}{\al}}{v} \simeq \tur{n-\left\lceil \frac{n}{\al}\right\rceil}{\al -1}$. By Lemma~\ref{lem_main1}, we obtain
\[
\fT(n,\al) = \fT(n-1,\al) +  \fT(n-\left\lceil \frac{n}{\al}\right\rceil,\al -1). \qedhere
\]
\end{proof}

A consequence of Lemma~\ref{lem_ind} is that $\fT(n-1,\al) < \fT(n,\al)$. Indeed, the cases $\al = 1$ and $\al = n$ are trivial, and the term $\fT(n-\left\lceil \frac{n}{\al}\right\rceil,\al -1)$ is always strictly positive when $2 \le \al \le n-1$. 

\begin{cor} \label{cor_f1inc}
The function $\fT(n,\al)$ is strictly increasing in $n$ when $\al$ is fixed.
\end{cor}

We now state the upper bound on the Fibonacci index of graphs in the class ${\cal G}(n,\al)$.

\begin{thm} \label{thm_genUB}
Let $G$ be a graph of order $n$ with a stability number $\al$, then
\[
\F(G) \le \fT(n,\al),
\]
with equality if and only if $G \simeq \tur{n}{\al}$.
\end{thm}

\begin{proof}
The cases $\al = 1$ and $\al = n$ are straightforward. Indeed $G \simeq \tur{n}{1}$ when $\al = 1$, and $G \simeq \tur{n}{n}$ when $\al = n$. We can assume that $2 \le \al \le n-1$, and thus $n \geq 3$. We now prove by induction on $n$ that if $G$ is extremal, then it is isomorphic to $\tur{n}{\al}$. 

The graph $G$ is $\al$-critical. Otherwise, there exists an edge $e \in E(G)$ such that $\al(G) = \al(\gme{G}{e})$, and by Lemma \ref{lem_main1}, $F(G) < F(\gme{G}{e})$. This is a contradiction with $G$ being extremal.

Let us compute $F(G)$ thanks to Lemma~\ref{lem_main1}. Let $v \in V(G)$ of maximum degree $\Delta$. The vertex $v$ is not isolated since $\al < n$. Thus by Lemma~\ref{lem_crit}, $\al(\gp{G}{v})  =  \al$ and $\al(\gpp{G}{v}) = \al - 1$. On the other hand, If $\chi$ is the chromatic number of $G$, it is well-known that $n \le \chi \ . \ \al$ (see, e.g., Berge~\cite{Berge01}), and that $\chi \le \Delta + 1$ (see Brooks~\cite{Brooks41}). It follows that 
\begin{equation} \label{eq_ngpp} 
n(\gpp{G}{v}) = n - \Delta - 1 \leq n - \left\lceil \frac{n}{\al}\right\rceil.
\end{equation}
Note that $n(\gpp{G}{v}) \geq 1$ since $\al \geq 2$. 

We can apply the induction hypothesis on the graphs $\gp{G}{v}$ and $\gpp{G}{v}$. We obtain

\[
\begin{array}{ r c l p{0.35 \textwidth}}
\fT(n,\al) & \le & \F(G) &\textrm{as $G$ is extremal},\\
& = & \F(\gp{G}{v}) + \F(\gpp{G}{v}), & \textrm{by Lemma~\ref{lem_main1},}\\ 
& \le & \fT(n(\gp{G}{v}),\al(\gp{G}{v})) + \fT(n(\gpp{G}{v}),\al(\gpp{G}{v})), & \textrm{by induction,}\\
& = & \fT(n-1,\al) + \fT(n-\Delta-1,\al-1), &\\
& \le &  \fT(n-1,\al) +  \fT(n-\left\lceil \frac{n}{\al}\right\rceil,\al -1), & \textrm{by Eq.~\eqref{eq_ngpp} and Corollary~\ref{cor_f1inc},}\\
& = & \fT(n,\al) & \textrm{by Lemma~\ref{lem_ind}}.
\end{array}
\]
Hence  equality holds everywhere. In particular, by induction, the graphs $\gp{G}{v}$, $\gpp{G}{v}$ are extremal, and $\gp{G}{v} \simeq \tur{n-1}{\al}$, $\gpp{G}{v} \simeq \tur{n-\left\lceil \frac{n}{\al}\right\rceil}{\al -1}$. Coming back to $G$ from $\gp{G}{v}$ and $\gpp{G}{v}$ and recalling that $v$ has maximum degree, it follows that $G \simeq \tur{n}{\al}$.
\end{proof}

Corollary~\ref{cor_f1inc} states that $\fT(n,\al)$ is increasing in $n$. It was an easy consequence of Lemma~\ref{lem_ind}. The function $\fT(n,\al)$ is also increasing in $\al$. Theorem~\ref{thm_genUB} can be used to prove this fact easily as shown now.

\begin{cor} \label{cor_f1inc2}
The function $\fT(n,\al)$ is strictly increasing in $\al$ when $n$ is fixed.
\end{cor}

\begin{proof}
Suppose $2 \le \al \le n-1$. By Lemma~\ref{lem_main2} it is clear that $\fT(n,1) < \fT(n,\al) < \fT(n,n)$. Now, let $e$ be an edge of $\tur{n}{\al}$. Clearly $\al(\gme{\tur{n}{\al}}{e}) = \al+1$. Moreover, by Lemma~\ref{lem_main1} and Theorem~\ref{thm_genUB},
\[
\F(\tur{n}{\al}) < \F(\gme{\tur{n}{\al}}{e}) < \F(\tur{n}{\al+1}).
\]
Therefore, $\fT(n,\al) < \fT(n,\al+1)$.
\end{proof}

\section{Connected graphs}  \label{sec_conn}

We now consider graphs with maximum Fibonacci index inside the class ${\cal C}(n,\al)$. Such graphs are called \emph{extremal}. If $G$ is connected, the bound of Theorem \ref{thm_genUB} is clearly not tight, except when $\al = 1$, that is, when $G$ is a complete graph. We are going to prove that there is one extremal graph up to isomorphism, the Tur\'an-connected graph $\turc{n}{\al}$, with the exception of the cycle $\C{5}$ (see Theorem~\ref{thm_conUB}).  First, we need preliminary results and definitions to prove this theorem.

We denote by $\fTC(n,\al)$ the Fibonacci index of the Tur\'an-connected graph $\turc{n}{\al}$. An inductive formula for its value is given in the next lemma.

\begin{lem} \label{lem_f2}
Let $n$ and $\al$ be integers such that $1 \le \al \le n-1$. Then
\[
\fTC(n,\al) = \left\{ 
\begin{array}{l l l}
n+1 & \textrm{if} & \al = 1,\\
2^{n-1} + 1 & \textrm{if} & \al = n-1,\\
\fT(n-1,\al) + \fT(n',\al') & \textrm{if} & 2 \le \al \le n-2,\\
\end{array}
\right.
\]
where $n' = n-\left\lceil\frac{n}{\al}\right\rceil-\al+1$ and $\al' = \min(n',\al - 1)$.
\end{lem}
\begin{proof}
The cases $\al = 1$ and $\al = n-1$ are trivial by Lemma~\ref{lem_main2}. Suppose now that $2 \le \al \le n-2$. Let $v$ be a vertex of maximum degree in $\turc{n}{\al}$. We apply Lemma~\ref{lem_main1} to compute $F(\turc{n}{\al})$. Observe that the graphs $\gp{\turc{n}{\al}}{v}$ and $\gpp{\turc{n}{\al}}{v}$ are both Tur\'an graphs when $2 \le \al \le n-2$. 

The graph $\gp{\turc{n}{\al}}{v}$ is isomorphic to $\tur{n-1}{\al}$. Let us show that  $\gpp{\turc{n}{\al}}{v}$ is isomorphic to $\tur{n'}{\al'}$. By definition of a Tur\'an-connected graph, $d(v)$ is equal to $\left\lceil\frac{n}{\al}\right\rceil+\al-2$. Thus 
\[
n(\gpp{\turc{n}{\al}}{v}) = n - d(v) - 1 = n'.
\]
If $\al < \frac{n}{2}$, then $\turc{n}{\al}$ has a clique of order at least 3 and $\al(\gpp{\turc{n}{\al}}{v}) = \al - 1 \leq n'$. Otherwise, $\gpp{\turc{n}{\al}}{v} \simeq \overline{\K{n'}}$ and $\al(\gpp{\turc{n}{\al}}{v}) = n' \leq \al - 1$. Therefore $\al(\gpp{\turc{n}{\al}}{v}) =  \min(n',\al - 1)$ in both cases.

By Lemma~\ref{lem_main1}, these observations leads to 
\[
\fTC(n,\al) =  \fT(n-1,\al) + \fT(n',\al'). \qedhere
\]
\end{proof}

\begin{defn}
A \emph{bridge} in a connected graph $G$ is an edge $e \in E(G)$ such that the graph $\gme{G}{e}$ is no more connected. To a bridge $e = v_1 v_2$ of $G$ which is $\al$-safe, we associate a \emph{decomposition} $\Dec{G_1}{v_1}{G_2}{v_2}$ such that $v_1 \in V(G_1)$, $v_2 \in V(G_2)$, and $G_1, G_2$ are the two connected components of $\gme{G}{e}$. A decomposition is said to be $\al$-\emph{critical} if $G_1$ is $\al$-critical.
\end{defn}

\begin{lem} \label{lem_compCrit}
Let $G$ be a connected graph. If $G$ is extremal, then either $G$ is $\al$-critical or $G$ has an $\al$-critical decomposition.
\end{lem}
\begin{proof}
We suppose that $G$ is not $\al$-critical and we show that it must contain an $\al$-critical decomposition. 

Let $e$ be an $\al$-safe edge of $G$. Then $e$ must be a bridge. Otherwise, the graph $\gme{G}{e}$ is connected, has the same order and stability number as $G$ and satisfies $\F(\gme{G}{e}) > \F(G)$ by Lemma~\ref{lem_main1}. This is a contradiction with $G$ being extremal. Therefore $G$ contains at least one $\al$-safe bridge defining a decomposition of $G$. 

Let us choose a decomposition $\Dec{G_1}{v_1}{G_2}{v_2}$ such that $G_1$ is of minimum order. Then, $G_1$ is $\al$-critical. Otherwise, $G_1$ contains an $\al$-safe bridge $e' = w_1 w_2$, since the edges of $G$ are $\al$-critical or $\al$-safe bridges by the first part of the proof. Let $\Dec{H_1}{w_1}{H_2}{w_2}$ be the decomposition of $G$ defined by $e'$, such that $v_1 \in V(H_2)$. Then $n(H_1) < n(G_1)$, which is a contradiction. Hence the decomposition $\Dec{G_1}{v_1}{G_2}{v_2}$ is $\al$-critical.
\end{proof}

\begin{thm} \label{thm_conUB}
Let $G$ be a connected graph of order $n$ with a stability number $\al$, then
\[
\F(G) \le \fTC(n,\al),
\]
with equality if and only if $G \simeq \turc{n}{\al}$ when $(n,\al) \neq (5,2)$, and $G \simeq \turc{5}{2}$ or $G \simeq \C{5}$ when $(n,\al) = (5,2)$.
\end{thm}

\begin{proof}
 We prove by induction on $n$ that if $G$ is extremal, then it is isomorphic to $\turc{n}{\al}$ or $\C{5}$. To handle more easily the general case of the induction (in a way to avoid the extremal graph $\C{5}$), we consider all connected graphs with up to 6 vertices as the basis of the induction. For these basic cases, we refer to the report of an exhaustive automated verification~\cite{GPHRepFiboAlpha}. We thus suppose that $n \ge 7$.

We know by Lemma~\ref{lem_compCrit} that either $G$ has an $\al$-critical decomposition or $G$ is $\al$-critical. We consider now these two situations.

\paragraph{1) $G$ has an $\al$-critical decomposition.} We prove in three steps that $G \simeq \turc{n}{\al}$: (\emph{i}) We establish that for every decomposition $\Dec{G_1}{v_1}{G_2}{v_2}$, the graph $G_i$ is extremal and is isomorphic to a Tur\'an-connected graph such that $d(v_i) = \Delta(G_i)$, for $i = 1,2$. (\emph{ii}) We show that if such a decomposition is $\al$-critical, then $G_1$ is a clique. (\emph{iii}) We prove that $G$ is itself isomorphic to a Tur\'an-connected graph.

(\emph{i}) For the first step, let $\Dec{G_1}{v_1}{G_2}{v_2}$ be a decomposition of $G$, $n_1$ be the order of $G_1$, and $\al_1$ its stability number. We prove that $G_1 \simeq \turc{n_1}{\al_1}$ such that $d(v_1) = \Delta(G_1)$. The argument is identical for $G_2$. By Lemma~\ref{lem_main1}, we have
\[
 \F(G) = \F(G_1) \F(\gp{G_2}{v_2}) + \F(\gp{G_1}{v_1}) \F(\gpp{G_2}{v_2}).
\]
By the induction hypothesis, $\F(G_1) \leq \fTC(n_1,\al_1)$. The graph $\gp{G_1}{v_1}$ has an order $n_1 - 1$ and a stability number $\leq \al_1$. Hence by Theorem~\ref{thm_genUB} and Corollary~\ref{cor_f1inc2}, $ \F(\gp{G_1}{v_1}) \leq \fT(n_1-1,\al_1)$. It follows that
\begin{equation} \label{eq_con}
 \F(G)  \le \fTC(n_1,\al_1) \F(\gp{G_2}{v_2}) + \fT(n_1 - 1,\al_1) \F(\gpp{G_2}{v_2}).
\end{equation}
As $G$ is supposed to be extremal, equality occurs. It means that $\gp{G_1}{v_1} \simeq \tur{n_1 - 1}{\al_1}$ and $G_1$ is extremal. If $G_1$ is isomorphic to $\C{5}$, then $n_1 = 5$, $\al_1=2$ and $\F(G_1) = \fTC(5,2)$.  However, $\F(\gp{G_1}{v_1}) =  \F(\Path{4}) < \fT(4,2)$. By~\eqref{eq_con}, this leads to a contradiction with $G$ being extremal. Thus, $G_1$ must be isomorphic to $\turc{n_1}{\al_1}$. Moreover, $v_1$ is a vertex of maximum degree of $G_1$. Otherwise, $\gp{G_1}{v_1}$ cannot be isomorphic to the graph $\tur{n_1 - 1}{\al_1}$.

(\emph{ii}) The second step is easy. Let $\Dec{G_1}{v_1}{G_2}{v_2}$ be an $\al$-critical decomposition of $G$, that is, $G_1$ is $\al$-critical. By (\emph{i}), $G_1$ is isomorphic to a Tur\'an-connected graph. The complete graph is the only Tur\'an-connected graph which is $\al$-critical. Therefore, $G_1$ is a clique. 

(\emph{iii}) We now suppose that $G$ has an $\al$-critical decomposition $\Dec{G_1}{v_1}{G_2}{v_2}$ and we show that $G \simeq \turc{n}{\al}$. Let $n_1$ be the order of $G_1$ and $\al_1$ its stability number. As $v_1 v_2$ is an $\al$-safe bridge, it is clear that $n(G_2) = n - n_1$ and $\al(G_2) = \al - \al_1$. By (\emph{i}) and (\emph{ii}), $G_1$ is a clique (and thus $\al_1 = 1$), $G_2 \simeq\turc{n - n_1}{\al - 1}$, and $v_2$ is a vertex of maximum degree in $G_2$.  

If $\al = 2$, then $G_2$ is also a clique in $G$. By Lemma~\ref{lem_main1} and the fact that $\F(\K{n}) = n+1$ we have,
\[
\begin{array}{r c l}
\F(G) & = & \F(\gp{G}{v_1}) + \F(\gpp{G}{v_1}),\\
 & = & n_1 (n - n_1 + 1) + (n-n_1) =  n + n \ n_1 - n_1^2.
\end{array}
\]
When $n$ is fixed, this function is maximized when $n_1 = \frac{n}{2}$. That is, when $G_1$ and $G_2$ are balanced cliques. This appears if and only if $G \simeq \turc{n}{2}$. 

Thus we suppose that $\al \geq 3$. In other words, $G$ contains at least three cliques: the clique $G_1$ of order $n_1$; the clique $H$ containing $v_2$ and a clique $H'$ in $G_2$ linked to $H$ by an $\al$-safe bridge $v_2 v_3$. Let $k =  \frac{n-n_1}{\al-1}$, then the order of $H$ is $\left\lceil k \right\rceil$ and the order of $H'$ is $\left\lceil k \right\rceil$ or $\left\lfloor k \right\rfloor$ (recall that $G_2 \simeq \turc{n-n_1}{\al -1}$). These cliques are represented in Figure~\ref{fig_case1}. 

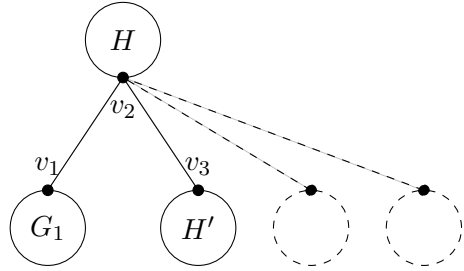
\begin{figure}[!ht]
\begin{center}
\begin{tikzpicture}
\draw[fill=black] (0,0) circle (2pt) -- (1,1.5);
\draw (0,0) node[above=2pt] {$v_1$};
\draw (0,-0.5) circle (0.5cm);
\draw (0,-0.5) node {$G_1$};
\draw[fill=black] (1,1.5) circle (2pt);
\draw (1,1.5) node[below=5pt] {$v_2$};
\draw (1,2) circle (0.5cm);
\draw (1,2) node {$H$};
\draw[fill=black] (2,0) circle (2pt) -- (1,1.5);
\draw (2,0) node[above=2pt] {$v_3$};
\draw (2,-0.5) circle (0.5cm);
\draw (2,-0.5) node {$H'$};
\draw[dashed, fill=black] (3.5,0) circle (2pt) -- (1,1.5);
\draw[dashed] (3.5,-0.5) circle (0.5cm);
\draw[dashed, fill=black] (5,0) circle (2pt) -- (1,1.5);
\draw[dashed] (5,-0.5) circle (0.5cm);
\end{tikzpicture}
\caption{Cliques in the graph $G$} \label{fig_case1}
\end{center}
\end{figure}

To prove that $G$ is isomorphic to a Tur\'an-connected graph, it remains to show that the clique $G_1$ is balanced with the cliques $H$ and $H'$. We consider the decomposition defined by the $\al$-safe bridge $v_2 v_3$. By~(\emph{i}), $G_1$ and $H$ are cliques of a Tur\'an-connected graph, and $H$ is a clique with maximum order in this graph (recall that $v_2$ is a vertex of maximum degree in $G_2$). Therefore $\left\lceil k \right\rceil -1 \le n_1 \le  \left\lceil k \right\rceil$, showing that $G_1$ is balanced with $H$ and $H'$.

\paragraph{2) $G$ is $\al$-critical.}  Under this hypothesis, we prove that $G$ is a complete graph, and thus is isomorphic to a Tur\'an-connected graph.

Suppose that $G$ is not complete. Let $v$ be a vertex of $G$ with a maximum degree $d(v) = \Delta$. As $G$ is connected and $\al$-critical, the graph $\gp{G}{v}$ is connected by Lemma~\ref{lem_crit1}. By Lemma~\ref{lem_crit}, $\al(\gp{G}{v}) = \al$ and $\al(\gpp{G}{v}) = \al - 1$. Moreover, $n(\gp{G}{v}) = n-1$ and $n(\gpp{G}{v}) = n - \Delta - 1$. By the induction hypothesis and Theorem~\ref{thm_genUB}, we get
\[
 \F(G) = \F(\gp{G}{v}) + \F(\gpp{G}{v}) \le \fTC(n-1,\al) + \fT(n-\Delta-1,\al-1).
\] 
Therefore, $G$ is extremal if and only if $\gpp{G}{v} \simeq \tur{n-\Delta-1}{\al-1}$ and $\gp{G}{v}$ is extremal. However, $\gp{G}{v}$ is not isomorphic to $\C{5}$ as $n \ge 7$. Thus $\gp{G}{v} \simeq \turc{n-1}{\al}$.

So, the graph $G$ is composed by the graph $\gp{G}{v} \simeq \turc{n-1}{\al}$ and an additional vertex $v$ connected to $\turc{n-1}{\al}$ by $\Delta$ edges. 

There must be an edge between $v$ and a vertex $v'$ of maximum degree in $\gp{G}{v}$, otherwise $\gpp{G}{v}$ is not isomorphic to a Tur\'an graph. The vertex $v'$ is adjacent to $\left\lceil \frac{n-1}{\al} \right\rceil + \al - 2$ vertices in $\gp{G}{v}$ and it is adjacent to $v$, that is,
\[
d(v') = \left\lceil \frac{n-1}{\al} \right\rceil + \al - 1.
\]
It follows that 
\begin{equation} \label{eq_del}
\Delta \ge d(v') >  \left\lceil \frac{n-1}{\al} \right\rceil
\end{equation}
as $G$ is not a complete graph. 

On the other hand, $v$ is adjacent to each vertex of some clique $H$ of $\gp{G}{v}$ since $\gpp{G}{v}$ has a stability number $\al - 1$. As this clique has order at most $\left\lceil \frac{n-1}{\al} \right\rceil$, $v$ must be adjacent to a vertex $w \notin H$ by \eqref{eq_del}. 

We observe that the edge $vw$ is $\al$-safe. This is impossible as $G$ is $\al$-critical. It follows that $G$ is a complete graph and the proof is completed. \qedhere
\end{proof}

The study of the maximum Fibonacci index inside the class ${\cal T}(n,\al)$ of trees with order $n$ and stability number $\al$ is strongly related to the study done in this section for the class ${\cal C}(n,\al)$. Indeed, due to the fact that trees are bipartite, a tree in ${\cal T}(n,\al)$ has always a stability number $\al \geq \frac{n}{2}$. Moreover, the Tur\'an-connected graph $\turc{n}{\al}$ is a tree when $\al \ge \frac{n}{2}$. Therefore, the upper bound on the Fibonacci index for connected graphs is also valid for trees. We thus get the next corollary with in addition the exact value of $\fTC(n,\al)$.

\begin{cor} Let $G$ be a tree of order $n$ with a stability number $\al$, then
\[
\F(G) \le 3^{n-\al-1} 2^{2 \al - n + 1}+2^{n-\al-1},
\]
with equality if and only if $G \simeq \turc{n}{\al}$.
\end{cor}

\begin{proof}
It remains to compute the exact value of $\fTC(n,\al)$. When $\al \ge \frac{n}{2}$, the graph $\turc{n}{\al}$ is composed by one central vertex $v$ of degree $\al$ and $\al$ pending paths of length 1 or 2 attached to $v$. An extremity of a pending path of length 2 is a vertex $w$ such that $w \notin  \mathcal{N}(v)$. Thus there are $x = n-\al-1$ pending paths of length 2 since $\mathcal{N}(v)$ has size $\al +1$, and there are $y = \al - x = 2 \al - n + 1$ pending paths of length 1. We apply Lemma~\ref{lem_main1} on $v$ to get
\[
\fTC(n,\al) = \F(\K{2})^x \F(\K{1})^y + \F(\K{1})^x = 3^x 2^y+2^x. \qedhere
\] 
\end{proof}

\section{Observations} \label{sec_conc}

Tur\'an graphs $\tur{n}{\al}$ have minimum size inside ${\cal G}(n,\al)$ by the Theorem of Tur\'an~\cite{Turan}. Christophe et al.~\cite{GphStableMax} give a tight lower bound for the connected case of this theorem, and Bougard and Joret~\cite{Bougard08} characterized the extremal graphs, which happen to contain the $\turc{n}{\al}$ graphs as a subclass.

By these results and Theorems \ref{thm_genUB} and \ref{thm_conUB}, we can observe the following relations between graphs with minimum size and maximum Fibonacci index. The graphs inside ${\cal G}(n,\al)$ minimizing $m(G)$ are exactly those which maximize $\F(G)$. This is also true for the graphs inside ${\cal C}(n,\al)$, except that there exist other graphs with minimum size than the Tur\'an-connected graphs. 

However, these observations are not a trivial consequence of the fact that $\F(G) < \F(\gme{G}{e})$ where $e$ is any edge of a graph $G$. As indicated in our proofs, the latter property only implies that a graph maximizing $\F(G)$ contains only $\al$-critical edges (and $\al$-safe bridges for the connected case). Our proofs use a deep study of the structure of the extremal graphs to obtain Theorems \ref{thm_genUB} and \ref{thm_conUB}.

We now give additional examples showing that the intuition that more edges imply fewer stable sets is wrong. Pedersen and Vestergaard~\cite{PV06} give the following example. Let $r$ be an integer such that $r\ge 3$, $G_1$ be the Tur\'an graph $\tur{2r}{r}$ and $G_2$ be the star $\Star{2r}$. The graphs  $G_1$ and $G_2$ have the same order but $G_1$ has less edges ($r$) than $G_2$ ($2r-1$). Nevertheless, observe that $F(G_1) = 3^{r} < F(G_2) = 2^{2r-1}+1$. This example does not take into account the stability number since $\al(G_1) = r$ and $\al(G_2) = 2r-1$. 

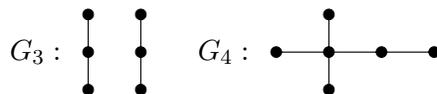
\begin{figure}[!ht]
\begin{center}
\begin{tikzpicture}
\draw (-0.7,1) node {$G_3:$};
\draw[fill=black] (0,0.5) circle (2pt) -- (0,1) circle (2pt) -- (0,1.5) circle (2pt);
\draw[fill=black] (0.7,0.5) circle (2pt) -- (0.7,1) circle (2pt) -- (0.7,1.5) circle (2pt);
\draw (1.8,1) node {$G_4:$};
\draw[fill=black] (2.5,1) circle (2pt) -- (3.2,1);
\draw[fill=black] (3.2,0.5) circle (2pt) -- (3.2,1) circle (2pt) -- (3.2,1.5) circle (2pt);
\draw[fill=black] (3.2,1) -- (3.9,1) circle (2pt) -- (4.6,1) circle (2pt);
\end{tikzpicture}
\caption{Graphs with same order and stability number} \label{fig_cexa}
\end{center}
\end{figure}

We propose a similar example of pairs of graphs with the same order and the same stability number (see the graphs $G_3$ and $G_4$ on Figure~\ref{fig_cexa}). These two graphs are inside the class ${\cal G}(6,4)$, however $m(G_3) < m(G_4)$ and $\F(G_3) < \F(G_4)$. Notice that we can get such examples inside ${\cal G}(n,\al)$ with $n$ arbitrarily large, by considering the union of several disjoint copies of $G_3$ and $G_4$.

These remarks and our results suggest some questions about the relations between the size, the stability number and the Fibonacci index of graphs. What are the lower and upper bounds for the Fibonacci index inside the class ${\cal G}(n,m)$ of graphs order $n$ and size $m$; or inside the class ${\cal G}(n,m,\al)$ of graphs order $n$, size $m$ and stability number $\al$? Are there classes of graphs for which more edges always imply fewer stable sets? We think that these questions deserve to be studied.

\section*{Acknowledgments}

The authors thank Gwena\"el Joret for helpful suggestions.


\begin{thebibliography}{10}

\bibitem{Berge01}
{\sc Berge, C.}
\newblock {\em {The Theory of Graphs}}.
\newblock Dover Publications, New York, 2001.

\bibitem{Bougard08}
{\sc Bougard, N., and Joret, G.}
\newblock {Tur\'an Theorem and $k$-connected graphs}.
\newblock Accepted for publication in \emph{J. Graph Theory} (2008), 13 pages.

\bibitem{Brooks41}
{\sc Brooks, {\protect R.-L}.}
\newblock {On colouring the nodes of a network}.
\newblock {\em Proc. Cambridge Philos. Soc. 37\/} (1941), 194 -- 197.

\bibitem{Byskov04}
{\sc Byskov, {\protect J.M}.}
\newblock {Enumerating maximal independent sets with applications to graph
  colouring}.
\newblock {\em Oper. Res. Lett. 32\/} (2004), 547 -- 556.

\bibitem{GphStableMax}
{\sc Christophe, J., Dewez, S., Doignon, {\protect J.-P}., Elloumi, S.,
  Fasbender, G., Gr\'egoire, P., Huygens, D., Labb\'e, M., M\'elot, H., and
  Yaman, H.}
\newblock { Linear inequalities among graph invariants: using GraPHedron to
  uncover optimal relationships}.
\newblock Accepted for publication in \emph{Networks} (2008), 24 pages.

\bibitem{ErdG61}
{\sc Erd\"os, P., and Gallai, T.}
\newblock {On the minimal number of vertices representing the edges of a
  graph.}
\newblock {\em Magyar Tud. Akad. Mat. Kutat\'o Int. K\"ozl. 6\/} (1961),
  181--203.

\bibitem{GPHRepFiboAlpha}
{\sc {\protect GraPHedron}}.
\newblock {Reports on the study of the Fibonacci index and the stability number
  of graphs and connected graphs}.
\newblock URL: {\tt www.graphedron.net/index.php?page=viewBib\&bib=7}.

\bibitem{GP86}
{\sc Gutman, I., and Polansky, {\protect O.E}.}
\newblock {\em {Mathematical Concepts in Organic Chemistry}}.
\newblock Springer-Verlag, Berlin, 1986.

\bibitem{HW07}
{\sc Heuberger, C., and Wagner, S.}
\newblock {Maximizing the number of independent subsets over trees with bounded
  degree}.
\newblock Accepted for publication in \emph{J. Graph Theory} (2008), 14 pages.

\bibitem{Joret07}
{\sc Joret, G.}
\newblock {\em {Entropy and Stability in Graphs}}.
\newblock PhD thesis, Universit\'e Libre de Bruxelles, Belgium, 2007.

\bibitem{KTWZ07}
{\sc Knopfmacher, A., Tichy, {\protect R.F.}., Wagner, S., and Ziegler, V.}
\newblock {Graphs, partitions and Fibonacci numbers}.
\newblock {\em Discrete Appl. Math. 155\/} (2007), 1175 -- 1187.

\bibitem{LLW03}
{\sc Li, X., Li, Z., and Wang, L.}
\newblock {The Inverse Problems for Some Topological Indices in Combinatorial
  Chemistry}.
\newblock {\em J. Comput. Biol. 10}, 1 (2003), 47 -- 55.

\bibitem{LZG05}
{\sc Li, X., Zhao, H., and Gutman, I.}
\newblock {On the Merrifield-Simmons Index of Trees}.
\newblock {\em MATCH Comm. Math. Comp. Chem. 54\/} (2005), 389 -- 402.

\bibitem{Lovasz86}
{\sc Lov\'asz, L., and Plummer, {\protect M.D}.}
\newblock {\em {Matching Theory}}.
\newblock Akad\'emiai Kiad\'o - North-Holland, Budapest, 1986.

\bibitem{GphDesc}
{\sc M\'elot, H.}
\newblock Facet defining inequalities among graph invariants: the system
  {GraPHedron}.
\newblock Accepted for publication in \emph{Discrete Appl. Math.} (2007), 17
  pages.

\bibitem{MS89}
{\sc Merrifield, {\protect R.E.}., and Simmons, {\protect H.E.}.}
\newblock {\em {Topological Methods in Chemistry}}.
\newblock Wiley, New York, 1989.

\bibitem{PV05}
{\sc Pedersen, {\protect A.S}., and Vestergaard, {\protect P.D}.}
\newblock The number of independent sets in unicyclic graphs.
\newblock {\em Discrete Appl. Math. 152\/} (2005), 246 -- 256.

\bibitem{PV06}
{\sc Pedersen, {\protect A.S}., and Vestergaard, {\protect P.D}.}
\newblock {Bounds on the Number of Vertex Independent Sets in a Graph}.
\newblock {\em Taiwanese J. Math. 10}, 6 (2006), 1575 -- 1587.

\bibitem{PV}
{\sc Pedersen, {\protect A.S}., and Vestergaard, {\protect P.D}.}
\newblock {An Upper Bound on the Number of Independent Sets in a Tree}.
\newblock {\em Ars Combin. 84\/} (2007), 85 -- 96.

\bibitem{PT82}
{\sc Prodinger, H., and Tichy, {\protect R.F.}.}
\newblock Fibonacci numbers of graphs.
\newblock {\em Fibonacci Quart. 20}, 1 (1982), 16 -- 21.

\bibitem{TW05}
{\sc Tichy, {\protect R.F}., and Wagner, S.}
\newblock {Extremal Problems for Topological Indices in Combinatorial
  Chemistry}.
\newblock {\em J. Comput. Biol. 12}, 7 (2005), 1004 -- 1013.

\bibitem{Turan}
{\sc Tur{\'a}n, P.}
\newblock Eine {E}xtremalaufgabe aus der {G}raphentheorie.
\newblock {\em Mat. Fiz. Lapok 48\/} (1941), 436--452.

\bibitem{Wag07}
{\sc Wagner, S.}
\newblock {Extremal trees with respect to Hosoya Index and Merrifield-Simmons
  Index}.
\newblock {\em MATCH Comm. Math. Comp. Chem. 57\/} (2007), 221 -- 233.

\bibitem{WH06}
{\sc Wang, H., and Hua, H.}
\newblock {Unicycle graphs with extremal Merrifield-Simmons Index}.
\newblock {\em J. Math. Chem. 43}, 1 (2008), 202 -- 209.

\bibitem{WHW07}
{\sc Wang, M., Hua, H., and Wang, D.}
\newblock {The first and second largest Merrifield-Simmons indices of trees
  with prescribed pendent vertices}.
\newblock {\em J. Math. Chem. 43}, 2 (2008), 727 -- 736.

\end{thebibliography}
\end{document}